\documentclass[epjc3,twocolumn]{svjour3}          %

\RequirePackage{amsmath}
\RequirePackage{amssymb}
\RequirePackage[T1]{fontenc}

\RequirePackage{graphicx}
\RequirePackage{mathptmx}      
\RequirePackage{flushend}
\RequirePackage[numbers,sort&compress]{natbib}
\RequirePackage[colorlinks,citecolor=blue,urlcolor=blue,linkcolor=blue]{hyperref}

\journalname{Eur. Phys. J. C}
\usepackage{ulem}
\RequirePackage{color}

\begin{document}

\def\nat{Nature}
\def\prl{Phys. Rev. Lett.}
\def\prb{Phys. Rev. B}
\def\prc{Phys. Rev. C}
\def\prd{Phys. Rev. D}

\def\mnras{Mon. Not. Roy. Astr. Soc.}
\def\apj{Astrophys. J.}
\def\apjl{Astrophys. J. Lett.}
\def\apjs{Astrophys. J. Suppl. Ser.}
\def\aa{Astron. Astrophys.}
\def\aap{Astron. Astrophys.}
\def\actaa{Acta Astronomica}
\def\aapr{Astron. Astrophys. Rev.}
\def\plb{Phys. Lett. B}
\def\pr{Phys. Rev.}
\def\araa{Annual Rev. of Astron. Astrophys.}

\def\pasj{Publications of the Astronomical Society of Japan }
\def\pasp{Publications of the Astronomical Society of the Pacific}
\def\zap{Zeitschrift f{\"u}r Astrophysik}

\def\npa{Nuclear Physics A}
\def\nphysa{Nucl. Phys.}
\def\physrep{Phys. Rep.}
\def\jcap{Journal of Cosmology and Astroparticle Physics}

\def\beq#1{\begin{equation}\label{#1}}
\def\eeq{\end{equation}}

\newcommand{\bear}[1]{\begin{eqnarray}\label{#1}}
\newcommand{\ear}{\end{eqnarray}}

\newcommand{\R}{{\mathbb R}}
\newcommand{\p}{\partial}
\newcommand{\nn}{\nonumber}

\title{Quasinormal modes in the field of a dyon-like dilatonic black hole}

\author{A.N.~Malybayev\thanksref{e1,addr1} 
        \and
        K.A.~Boshkayev\thanksref{e2,addr1,addr2,addr3} 
        \and
        V.D.~Ivashchuk\thanksref{e3,addr4,addr5}
}
\thankstext{e1}{e-mail: algis\_malybayev@mail.ru}
\thankstext{e2}{e-mail: kuantay@mail.ru}
\thankstext{e3}{e-mail: ivashchuk@mail.ru}

\institute{ Department of Theoretical and Nuclear Physics, Al-Farabi Kazakh National University,  \\
Al-Farabi ave., 71, Almaty 050040, Kazakhstan\label{addr1}
\and
National Nanotechnology Laboratory of Open Type,
Al-Farabi Kazakh National University, \\
Al-Farabi ave., 71, Almaty 050040, Kazakhstan\label{addr2}
\and
Department of Engineering Physics, Satbayev University, 22 Satbayev St., 050013 Almaty, Kazakhstan\label{addr3}
\and
Ratbay Myrzakulov Eurasian International Centre for Theoretical Physics, Tauelsizdik Ave 27/1-9, Nur-Sultan 010000, Kazakhstan\label{addr33}
\and
Center for Gravitation and Fundamental Metrology, VNIIMS,  \\           
46 Ozyornaya St., Moscow 119361,  Russian Federation\label{addr4}
\and
Institute of Gravitation and Cosmology, Peoples' Friendship University of Russia (RUDN University),\\ 6 Miklukho-Maklaya St., Moscow 117198, Russian Federation\label{addr5}
}

\date{Received: date / Accepted: date}

\maketitle

\begin{abstract}
Quasinormal modes of massless test scalar field in the background of gravitational field for a non-extremal dilatonic dyonic black hole are explored. The dyon-like black hole solution is considered in the gravitational $4d$ model involving two scalar fields and two 2-forms. It is governed by two 2-dimensional dilatonic coupling vectors $\vec{\lambda}_i$ obeying $\vec{\lambda}_i (\vec{\lambda}_1 + \vec{\lambda}_2) > 0$, $i =1,2$. The first law of black hole thermodynamics is given and the Smarr relation is verified.
Quasinormal modes for a massless  scalar (test) field in the eikonal approximation  are obtained and analysed. These modes depend upon a dimensionless parameter $a$ ($0 < a \leq 2$) which is a function of $\vec{\lambda}_i$. For limiting strong ($a = +0$) and weak ($a = 2$) coupling  cases,  they coincide with the well-known results for the Schwarzschild  and Reissner-Nordstr\"om solutions.  It is shown that the Hod conjecture, connecting the damping rate and the Hawking temperature, is satisfied  for $0 < a \leq 1$ and all allowed values of parameters. 
\end{abstract}


\section{Introduction}
The recent discovery/detection of gravitational waves  \cite{2016PhRvL.116f1102A,2019PhRvX...9c1040A,2020ApJ...892L...3A}  has strengthen a long-living interest to quasinormal modes (QNMs) \cite{1970Natur.227..936V,1971ApJ...170L.105P,1975RSPSA.344..441C,1984PhLA..100..231B,1984PhRvL..52.1361F,1984PhRvD..30..295F,1999LRR.....2....2K,1999CQGra..16R.159N,2009CQGra..26p3001B,2011RvMP...83..793K}, predicted by Vishveshwara in 1970. The detected gravitational waves were emitted during the final  (ringdown) stage of binary black hole mergers. The  frequencies of these waves were   governed by a certain  superpositions of  decaying oscillations, i.e. QNMs. The careful analysis of these experiments may be rather important since it can shed some light on nature of gravity in a strong field regime.

From the mathematical point of view, the quasinormal  mode (QNM) problem  can be  reduced to studying  the solutions to a wave equation  for  a scalar function  $\Phi(t,x)$ chosen in the following form
\begin{equation}
\label{0.0}
\Phi(t,x) = e^{-i\omega t} \Phi_{*} (x) ,
 \end{equation}
where $\Phi_{*}   = \Phi_{*}  (x)$ obeys  a Schr\"odinger-type equation 
\begin{equation}
\label{0.1}
\left(-\epsilon^2\frac{d^2}{dx^2} + V(x)\right) \Phi_{*}  = \omega^2 \Phi_{*},
\end{equation}
defined on a certain domain of real line $\mathbb{R}= (- \infty, + \infty)$, where $\epsilon > 0$ is  some parameter, e.g. $\epsilon = 1$; for reviews see \cite{1999LRR.....2....2K,1999CQGra..16R.159N,2009CQGra..26p3001B,2011RvMP...83..793K}. For  asymptotically flat black-hole solutions  
the functions $\Phi_{*}(x) $ are defined on  $\mathbb{R} $. In this case $x$ is   chosen as  a so-called tortoise coordinate (in the body of the paper denoted as $ R_{*} $), 
and (at least) for certain known spherically symmetric solutions 
(e.g. Schwarzschild and Reissner-Nordstr\"om ones) 
the potential $V(x)$ is a positively defined smooth function, having sufficiently fast fall off (to zero) in approaching either to the horizon ($x \to - \infty$) or to the spatial infinity $(x \to +\infty)$.  Usually QNM frequencies $\omega$  are defined as  complex numbers obeying ${\rm Re} \ \omega > 0$ and
${\rm Im} \ \omega < 0$, such that   the wave functions (\ref{0.0})  are exponentially damped in time as $(t \to +\infty)$,
corresponding to asymptotically  stable perturbations.  The QNM frequencies  
appear for the solutions to  equation (\ref{0.1}) which behave as outgoing waves at spatial infinity:
$\Phi_{*}(x)   \sim e^{ \frac{i \omega x}{\epsilon}}$ for $x \to \infty$
  (${\rm  Re} \ \omega > 0$) and ingoing ones at the horizon: 
 $\Phi_{*}(x)   \sim e^{- \frac{i\omega x}{\epsilon} }$ for $x \to-\infty$ with exponential
growth (in $|x|$) for $|\Phi_{*}(x)|$ as $|x| \to \infty$ (due to ${\rm Im} \ \omega < 0$).

For calculation of  QNM  \cite{2009CQGra..26p3001B,2011RvMP...83..793K} there exists a (most popular) 
method, introduced in Refs. \cite{ 1984PhLA..100..231B,1984PhRvL..52.1361F,1984PhRvD..30..295F},  which may be called as analytical  continuation method. The most transparent version of this method was recently proposed (and verified) in Ref. 
\cite{2020PhRvD.101b4008H}. Here we consider for simplicity the case of the  potential defined on all real axis $\mathbb{R}= (- \infty, + \infty)$ to avoid the boundary problems. (For more involved and subtle case when the  Schr\"odinger operator and effective potential are defined on $(0, + \infty)$ and proper boundary condition should be specified see   Ref. \cite{2021PhRvD.103d5001F}.)
The prescription is as follows: one should  start with the Schr\"odinger equation for non-relativistic quantum particle (of mass $1/2$) moving in the inverted potential $-V(x)$:  
\begin{equation}
\label{0.2}
 \left(-\hbar^2\frac{d^2}{dx^2} - V(x)\right) \Psi = E \Psi,
\end{equation}
where $\Psi = \Psi(x)$ is the wave function.
For a potential under consideration the inverted potential $- V(x)$ may have certain 
bounded states described by discrete spectrum energy levels $E_n = E(\hbar, n| -V)$, where
$n =  0, 1, \dots$.  The corresponding 
wave functions   $\Psi =  \Psi_n(x)$ for a proper potential under consideration is exponentially decaying at both infinities, i.e. 
    $\Psi_n \sim e^{\mp \sqrt{-E_n}\frac{x}{\hbar }  }$ as $ x \to \pm \infty $.
The  Hatsuda's  (analytical continuation) approach \cite{2020PhRvD.101b4008H} tells us that 
QNM frequencies for the potential $V(x)$ may be obtained from bounded states for the inverted potential $-V(x)$ by putting  formally  $\hbar = i \epsilon$ in (\ref{0.1}). Hence, due to this prescription 
we get the following QNM frequencies
\begin{equation}
\label{0.3}
\omega^2 = -E(\hbar = i\epsilon, n| -V),
\end{equation}
where $n = 0,1, \dots$ is called as (QNM) overtone number.

 It should be noted, that the method suggested in  
Ref.~\cite{2020PhRvD.101b4008H} for computation quasinormal frequencies  of spherically symmetric black holes, relates them to bound state energies of anharmonic oscillators  by using the analytic continuation in $\hbar$. It was stated in Ref.~\cite{2020PhRvD.101b4008H}, that the  known WKB  results  are easily reproduced by this method and, moreover, “the  perturbative  WKB  series  of  the  quasinormal  frequencies turn  out  to  be  Borel summable  divergent  series  both  for the  Schwarzschild  and  
for the Reissner-Nordstr\"om black holes.   

In this paper we continue our previous studies \cite{ABDI,ABI,AIMT} devoted to dilatonic dyon and dyon-like black hole solutions. Dilatonic and dyon-like black hole solutions were considered in numerous papers, see \cite{BronShikin,Gibbons,GM,GHS} and  \cite{ChHsuL,GKLTT,PTW,FIMS,GKO,Dav,GalZad}, respectively, and references therein. We note that earlier the main motivation for considering the dilaton scalar fields was coming from (super)string theory or certain higher-dimensional (e.g. supergravity) models.   

Here we study QNM spectrum in eikonal approximation for 
a special dyon-like dilatonic black hole solution  from Ref.~\cite{2020JPhCS1690a2143B}, 
with  electric and magnetic color charges $Q_1$ and $Q_2$, respectively, in the $4d$ model with metric $g$, two scalar fields $\varphi^1,\varphi^2$,  two 2-forms  $F^{(1)}$ and $F^{(2)}$, corresponding to two   dilatonic coupling vectors  $\vec{\lambda}_1$ and $\vec{\lambda}_2$  ($\vec{\lambda}_1 \neq - \vec{\lambda}_2$), respectively. 

These eikonal QNM  modes depend upon a dimensionless parameter $a$ ($0 < a \leq 2$) which is a function of $\vec{\lambda}_i$.
It should be noted that QNMs for dilatonic black holes were considered in numerous papers,
see Refs. \cite{Konoplya:2001ji,2001PhRvD..63f4009F,2002PhRvD..66h4007K,2005CQGra..22.1129C,2005JPhCS..24..123N,2013MPLA...2850109S,2015PhRvD..92f4022K,2018PhRvD..98j4042P,2019EPJC...79.1021B} and references therein.

The relation between the 2-forms and color charges are given by
\beq{0.01}
 F^{(1)} = Q_1  \tau_1, \qquad F^{(2)} = Q_2 \tau_2,
 \eeq
where $\tau_2 = {\rm vol}[S^2]$ is magnetic 2-form, 
which is volume form on 2-dimensional sphere and 
 $\tau_1$ is an ``electric'' 2-form.

We note that in the case of one scalar field $\varphi$ and two coupling constants 
$\lambda_1$, $\lambda_2$ the dyon-like ansatz 
was considered recently in Refs. \cite{ABI,AIMT,Dav,GalZad}. For  
  $\lambda_1 = \lambda_2 = \lambda$ our solutions from Ref. \cite{ABI} were dealing
  with a trivial non-composite generalization of 
dilatonic dyon black hole solutions in the model with one 2-form and one  scalar field  
which was considered in Ref.~\cite{ABDI}, see also Refs.~\cite{GKLTT,PTW,FIMS,GKO},  
 and references therein. 

The solutions with one scalar field from Refs. \cite{ABI,AIMT} may be  embedded
to the solutions under examination by considering the case of  collinear dilatonic coupling vectors: 
\beq{0.C}
  \vec{\lambda}_1 = \lambda_1 \vec{e}, \qquad \vec{\lambda}_2 = \lambda_2 \vec{e},
 \eeq
where $\vec{e}^2 = 1$, $\lambda_1 + \lambda_2 \neq 0$.

The paper is organised as follows: in section \ref{blackhole} we review the main properties of the black hole dyon solution from Ref.~\cite{2020JPhCS1690a2143B}. In section \ref{particularcase} we consider the physical parameters and particular cases of the dyonic black hole solutions.  In section \ref{qnms} we analytically derive the eikonal approximation for frequences of QNM corresponding to massless test scalar field in the background metric of our dyonic-like black hole solution and study their features. 
In section \ref{limitingcase2} we consider
two limiting cases $a = +0$ and $a = 2$, corresponding to the Schwarzschild 
and Reissner-Nordstr\"om black hole solutions. In section \ref{Hod} we test/check
the validity of the Hod conjecture \cite{Hod} for the solution under consideration 
when $0 < a \leq 2$. 
Finally, we summarize our conclusion in section \ref{conclusion}.

\section{Black hole dyon solutions}\label{blackhole}
The action of a model containing two scalar fields, 2-form and 
dilatonic coupling vectors which was considered in Ref.~\cite{2020JPhCS1690a2143B},  
is following
\bear{i.1}
 S= \frac{1}{16 \pi G}  \int d^4 x \sqrt{|g|}\biggl\{ {\cal R}[g] -
  g^{\mu \nu} \p_{\mu} \vec{\varphi}  \p_{\nu} \vec{\varphi}
 \qquad \qquad   \nn \\
 - \frac{1}{2} e^{2 \vec{\lambda}_1 \vec{\varphi}} F^{(1)}_{\mu \nu} F^{(1)\mu \nu }
 - \frac{1}{2} e^{2 \vec{\lambda}_2 \vec{\varphi}} F^{(2)}_{\mu \nu} F^{(2) \mu \nu}
 \biggr\},
\ear
where $g= g_{\mu \nu}(x)dx^{\mu} \otimes dx^{\nu}$ is the metric,  $|g| =   |\det (g_{\mu \nu})|$, 
 $\vec{\varphi} =  (\varphi^1,\varphi^2)$ is the vector of scalar fields 
 belonging to ${\R}^2$, 
 $F^{(i)} = dA^{(i)}  =  \frac{1}{2} F^{(i)}_{\mu \nu} dx^{\mu} \wedge dx^{\nu}$
is the $2$-form with $A^{(i)} = A^{(i)}_{\mu} dx^{\mu}$, $i =1,2$, 
$G$ is the gravitational constant,
 $\vec{\lambda}_1 = (\lambda_{1i}) \neq \vec{0}$, 
 $\vec{\lambda}_2 = (\lambda_{2i}) \neq \vec{0}$ are the dilatonic coupling  vectors  
 obeying 
 \beq{i.1a}
 \vec{\lambda_1} \neq - \vec{\lambda_2} 
 \eeq
and ${\cal R}[g]$ is the Ricci scalar. Here and in what follows we put $c =1$
(where $c$ is the speed of light in vacuum.)

We consider a  dyonic-like black hole
solution to the  field equations corresponding to the action
(\ref{i.1}) which 
has the following form \cite{2020JPhCS1690a2143B}
\bear{i4.9}
 ds^2 &=&   H^{a}\biggl\{ -  H^{-2a} \left( 1 - \frac{2\mu}{R} \right) dt^2
  +  \frac{dR^2}{1 - \frac{2\mu}{R}} + R^2  d \Omega^2 \biggr\},
 \\  \label{i4.10}
 \varphi^i &=& \nu^i \ln H , 
\ear
\vspace{-0.5cm}
\beq{i4.10em}
 F^{(1)} = \frac{Q_1}{H^2 R^2} dt \wedge dR, \qquad
  F^{(2)}  =  Q_2 \tau,    
\eeq
where $Q_1$ and $Q_2$ are (color) charges - electric and magnetic, respectively,
 $\mu > 0$ is the extremality parameter,
 $d \Omega^2 = d \theta^2 + \sin^2 \theta d \phi^2$
is the canonical metric on the unit sphere $S^2$  ($0< \theta < \pi$, $0< \phi < 2 \pi$),
 $\tau = \sin \theta d \theta \wedge d \phi$ is the standard volume form on $S^2$,  
   \beq{i4.7}
     H = 1 + \frac{P}{R},
   \eeq
   with $P > 0$ obeying
   \beq{i4.8a}
     P (P + 2 \mu) = \frac{1}{2} Q^2 .
   \eeq
All the rest parameters of the solution are defined as follows
\bear{i4.10a}
   a &=&  \frac{  ( \vec{\lambda}_1 + \vec{\lambda}_2 )^2}{ \Delta },
 \\  \label{i4.10n}
     \Delta &\equiv& \frac{1}{2}  (\vec{\lambda}_1 + \vec{\lambda}_2)^2 +
      \vec{\lambda}_1^2 \vec{\lambda}_2^2 - (\vec{\lambda}_1 \vec{\lambda}_2)^2,
\\ \label{i2.BB} 
\nu^i &=& \frac{ \lambda_{1i} \vec{\lambda}_2 ( \vec{\lambda}_1 + \vec{\lambda}_2 )
    - \lambda_{2i} \vec{\lambda}_1 ( \vec{\lambda}_1 + \vec{\lambda}_2 )}{ \Delta }, 
  \ear
$i = 1,2$ and
 \beq{i4.8c}
    Q_1^2  = \frac{ \vec{\lambda}_{2}( \vec{\lambda}_1 + \vec{\lambda}_2 )}{2 \Delta} Q^2, \qquad
    Q_2^2  = \frac{ \vec{\lambda}_{1}( \vec{\lambda}_1 + \vec{\lambda}_2 )}{2 \Delta} Q^2.
 \eeq

Here the following additional restrictions on dilatonic coupling  vectors are imposed      
     \beq{i4.8bdd}
        \vec{\lambda}_{i}( \vec{\lambda}_1 + \vec{\lambda}_2 ) > 0,  \qquad i = 1, 2. 
     \eeq
 
Correspondingly, we note that    
  \beq{i.18BD}
       \Delta > 0,
  \eeq
     is valid for   $\vec{\lambda}_1 \neq - \vec{\lambda}_2$.
     
   Due to relations (\ref{i4.8bdd}) and (\ref{i.18BD}) the $Q_s^2$ are well-defined. 
 Note that the restrictions  (\ref{i4.8bdd}) imply relations 
 $\vec{\lambda}_s \neq \vec{0}$, $s = 1,2$, and  (\ref{i.1a}).
   
     Indeed, in this case we have the sum of two non-negative terms in 
    (\ref{i2.BB}):  $(\vec{\lambda}_1 + \vec{\lambda}_2)^2 > 0$ and 
    \beq{i.18BC}
     C = \vec{\lambda}_1^2 \vec{\lambda}_2^2 - (\vec{\lambda}_1 \vec{\lambda}_2)^2 \geq 0,
    \eeq
  due to the Cauchy-Schwarz inequality. Moreover, $C = 0$ if and only if vectors 
  $\vec{\lambda}_1$ and $\vec{\lambda}_2$ are collinear. Relation 
   (\ref{i.18BC}) implies 
   \beq{i.18a}
     0 < a \leq 2.
   \eeq
   For non-collinear vectors  $\vec{\lambda}_1$ and $\vec{\lambda}_2$ 
   we get  $0 < a < 2$ while $a = 2$ for collinear ones.
    
  This solution may be verified just by a straightforward substitution into the
  equations of motion. 
                      
  The calculation of scalar curvature for the metric 
  $ds^2 = g_{\mu \nu} dx^{\mu} dx^{\nu}$ in (\ref{i4.9}) yields
    \beq{i4.8R}
      {\cal R}[g]  =  \frac{a(2-a)P^2 (R - 2 \mu)}{2 R^{3 - a} (R + P)^{2 + a}}.
    \eeq   

\section{Particular cases and physical parameters}\label{particularcase}

Here we analyze certain cases and physical parameters corresponding to the solutions under consideration.

\subsection{Non-collinear and collinear cases}
 
 {\bf Non-collinear case.} For  non-collinear vectors $\vec{\lambda}_1$ and $\vec{\lambda}_2$
 ($0 < a < 2$)  we obtain
 \beq{i4.9R}
          {\cal R}[g]  \to - \infty,
 \eeq  
 as $R \to + 0$ and hence we have a black hole with  a horizon at $R = 2 \mu$ and singularity
 at $R = +0$.  

{\bf Collinear case.} For  collinear vectors $\vec{\lambda}_1$,  $\vec{\lambda}_2$ from (\ref{0.C}) obeying
 $\vec{\lambda}_1 + \vec{\lambda}_2 \neq \vec{0}$ we obtain $\nu^i = 0$, $a = 2$ and
\beq{i4.11}
  Q_1^2 = \frac{\lambda_{2}}{\lambda_1 + \lambda_2} Q^2, \qquad
  Q_2^2 = \frac{ \lambda_{1}}{\lambda_1 + \lambda_2} Q^2,
\eeq
where $\lambda_1 \lambda_2 > 0$. 
By changing the radial variable, $R = r - P$, we get a little 
extension of the solution from Ref. \cite{ABI}
 \bear{i4.12}
  &&ds^2 =   - f(r) dt^2 +   f(r)^{-1} dr^2 + r^2  d \Omega^2_{2},
     \\    \label{i4.12FP}
  &&F^{(1)}= \frac{Q_1}{r^2} dt \wedge dr, \quad F^{(2)}  =  Q_2 \tau,
  \qquad   \vec{\varphi} = \vec{0},
    \ear
 where $f(r) = 1 - \frac{2GM}{r} + \frac{Q^2}{2r^2}$, $ Q^2 = Q_1^2 + Q_2^2$ and $GM = P + \mu$ =
 $\sqrt{\mu^2 + \frac{1}{2} Q^2}$.
 
 The metric in these variables coincides with the well-known Reissner-Nordstr\"om  
 metric governed by two  parameters: $GM > 0$ and $ Q^2 < 2 (GM)^2$. 
 We have two horizons in this case. The electric and magnetic charges are not independent but obey Eqs. (\ref{i4.11}).
 Note that to be consistent with the literature the net charge here is related to the charge of the Reissner-Nordstr\"om   black hole as follows $Q^2=2Q_{RN}^2$.

\subsection{Gravitational mass and scalar charges}

The definition of the ADM gravitational mass is obtained from Eq.~\eqref{i4.9} in the weak field regime by comparing with 
$g_{00} = - (1 - 2GM/R + O(1/R))$ 
\beq{i5.1}
 GM =   \mu + \frac{a}{2} P.
\eeq

In turn, the scalar charge vector $\vec{Q}_{\varphi} = ( Q_{\varphi}^1, Q_{\varphi}^2 )$ 
is derived  from (\ref{i4.10}) in the weak field regime using the following definition: $\varphi^i = Q_{\varphi}^i /R + O(1/R)$:
\beq{i5.1s}
 \vec{Q}_{\varphi} =    \vec{\nu} P,
\eeq
where $\vec{\nu}$ is given by Eq.~\eqref{i2.BB}.

By combining  relations (\ref{i5.1}) and (\ref{i5.1s}) we obtain the following identity
\begin{equation}
\label{i5.1id}
     2 (GM)^2   +    \vec{Q}_{\varphi}^2   = Q_1^2 + Q_2^2 + 2 \mu^2.
\end{equation}

This formula does not contain vectors $\vec{\lambda}_s$.

The identity (\ref{i5.1id}) may be verified 
by using  (\ref{i4.10a}), (\ref{i4.8c}) and the following  relation
\beq{i5.1simQ2}
   \vec{\nu}^2 = \frac{(\vec{\lambda}_1 + \vec{\lambda}_2)^2
    (\vec{\lambda}_1^2 \vec{\lambda}_2^2 - (\vec{\lambda}_1 \vec{\lambda}_2)^2 )}{ \Delta^2}
    = \frac{a (2-a)}{2}.
 \eeq

For further analyses it is convenient to introduce the following  dimensionless parameters 
\begin{equation}
\label{i5.1pq}
    p = P/\mu > 0, \qquad q = |Q|/(GM).
\end{equation}
We obtain
\begin{equation}
\label{i5.1fpq}
    f_*(p,a) = \frac{p(p+2)}{(1 + \frac{a}{2} p)^2} = \frac{q^2}{2} .
\end{equation}
The function $f_*(p,a)$ is monotonically increasing in $p$ on $(0, + \infty)$ for any $a \in (0, 2]$
since
\begin{equation}
\label{i5.1fpq2}
  \frac{\partial}{\partial p}  f_* = \frac{16 [(1 - \frac{a}{2} p) +1 ]}{(2 + a p)^3} > 0.
\end{equation}
Due to (\ref{i5.1fpq}) and $\lim_{p \to + \infty} f_*(p,a) = 4/a^2$ relation  
(\ref{i5.1fpq}) defines a one-to-one correspondence
between $p \in (0, +\infty)$ and $q \in (0, \frac{2 \sqrt{2}}{a})$ for any (fixed) $a \in (0, 2]$.
Thus, we have   
\begin{equation}
\label{i5.1qQ}
  0 < q^2 < \frac{8}{a^2}, \qquad 0 < Q^2 < \frac{8}{a^2} (GM)^2.  
\end{equation}
The inverse map $p(q) = p(q,a)$ is defined for any  $a \in (0, 2]$ 
as follows 
\begin{equation}
\label{i5.1qQ2}
  p = \frac{8 \sqrt{\frac{1}{2}(1-a)q^2 + 1} + 2 a q^2 -8}{8 - a^2 q^2}.  
\end{equation}

\subsection{Black hole thermodynamics}

In this subsection we consider black hole thermodynamics by calculating the Hawking temperature and entropy, checking the first law (of black hole thermodynamics) and testing the Smarr relation.

To this end, for simplicity here we put $\hbar = c = k_B = 1$.
The Bekenstein-Hawking (area) entropy $S = A/(4G)$,
associated with the black solution  (\ref{i4.7})  
at the horizon at $R = 2 \mu$, where $A$ is the horizon area,  reads
\beq{i5.2s}
 S = S_{BH} =   \frac{4 \pi \mu^2}{G}  \left(1 + \frac{P}{2 \mu}\right)^{a},
\eeq
while the related Hawking temperature  is following one
\beq{i5.2}
  T = T_H =  \frac{1}{8 \pi \mu} \left(1 + \frac{P}{2 \mu}\right)^{-a}.
\eeq

{ It may be verified that relations  (\ref{i5.1}), (\ref{i5.2s}) and (\ref{i5.2}) imply the first law of the black hole thermodynamics
\beq{i5.3t}
   dM = T dS + \Phi dQ
\eeq
as well as the Smarr formula
\beq{i5.3S}
   M = 2 T S + \Phi Q,
\eeq
where 
\beq{i5.3P}
    \Phi =   \frac{a Q}{4 G (P + 2 \mu) } .
\eeq

Relations (\ref{i5.3t}), (\ref{i5.3S}) may be presented in the following 
form 
\beq{i5.4t}
   dM = T dS + \Phi_1 dQ_1 + \Phi_2 dQ_2,
\eeq
\beq{i5.4S}
   M = 2 T S + \Phi_1 Q_1 + \Phi_2 Q_2,
\eeq
where 
\beq{i5.4P}
    \Phi_i =   \frac{ Q_i}{2 G (P + 2 \mu) },
\eeq
$i = 1,2$. In derivation of relations (\ref{i5.4t}), (\ref{i5.4S}) the following
identity is used
\beq{i5.4Q}
    Q_1^2 +  Q_2^2 = \frac{a}{2} Q^2.
\eeq

Let us clarify the physical sense of  potentials (\ref{i5.4P}). The first relation
in (\ref{i4.10em}) for  $F^{(1)} = d A^{(1)}$, has a special solution for $1$-form:
\beq{i5.5A1}
   A^{(1)} =   A^{(1)}_0 (R) dt = \frac{Q_1}{R + P} dt.
\eeq
Thus, we get 
\beq{i5.5AP1}
  \Phi_1 =  \frac{1}{2 G}  A^{(1)}_0 (2 \mu ),
\eeq
i.e. $\Phi_1$ is coinciding up to a factor $1/(2 G)$ with the value of the zero component of the first 
Abelian gauge field $A^{(1)}$, or electric potential, (in chosen gauge) for the field of electric charge 
at the horizon.

Now let us consider the magnetic term in (\ref{i4.10em}). The calculation of Hodge-dual
gives us 
\beq{i5.6}
  *F^{(2)}  =  \frac{Q_2}{H^{a} R^2} dt \wedge dR,    
\eeq
(here 
$*F_{\mu \nu} = \frac{1}{2} \sqrt{|g|} \varepsilon_{\mu \nu \rho \sigma} F^{\rho \sigma}$,
$\varepsilon_{0123} = 1$).
Relation (\ref{i2.BB}) implies 
\beq{i5.7}
    \vec{\lambda}_2 \vec{\nu} = \frac{a}{2} -1,     
\eeq
and hence (see (\ref{i4.10}))
\beq{i5.8}
 e^{2 \vec{\lambda}_2 \vec{\varphi}} = H^{a-2}.
\eeq
For $S$-dual  2-form   
\beq{i5.9}
 \Tilde{F}^{(2)} = d \Tilde{A}^{(2)} =  e^{2 \vec{\lambda}_2 \vec{\varphi}} *F^{(2)}
\eeq
we get  
\beq{i5.10F}
 \Tilde{F}^{(2)} = \frac{Q_2}{H^{2} R^2} dt \wedge dR
\eeq
and we can choose a corresponding $1$-form as follows
\beq{i5.10A}
 \Tilde{A}^{(2)} =    \Tilde{A}^{(2)}_0 (R) dt = \frac{Q_2}{R + P} dt.
\eeq
Hence, 
\beq{i5.11}
  \Phi_2 =  \frac{1}{2 G}  \Tilde{A}^{(2)}_0 (2 \mu ),
\eeq
i.e. $\Phi_2$ is coinciding up to a factor $1/(2 G)$ 
with the value of the zero component of the 
dual  Abelian gauge field $\Tilde{A}^{(2)}$, 
or dual electric potential, (in chosen gauge)  at the horizon, 
which corresponds to the field of magnetic charge  modulated by scalar fields.

\section{Quasinormal modes}\label{qnms}
\begin{figure*}[ht]
\centering
\begin{tabular}{lr}
\includegraphics[width=0.48\linewidth]{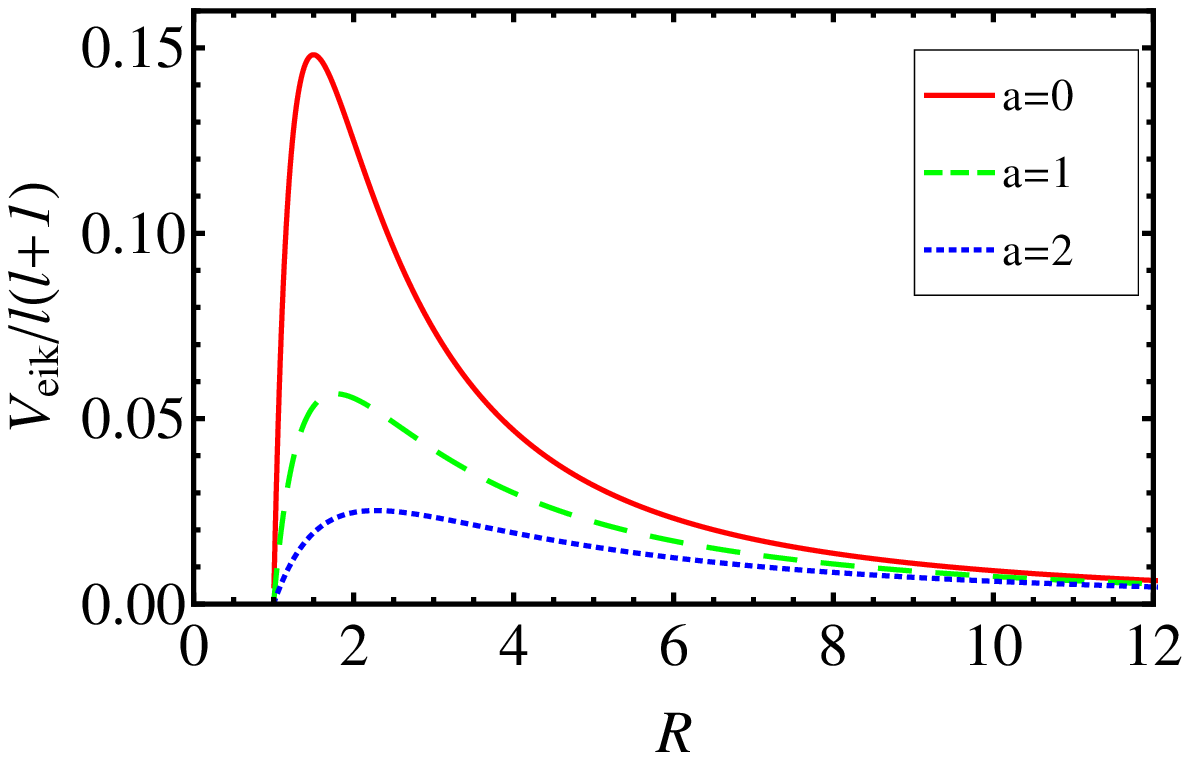} &\includegraphics[width=0.48\linewidth]{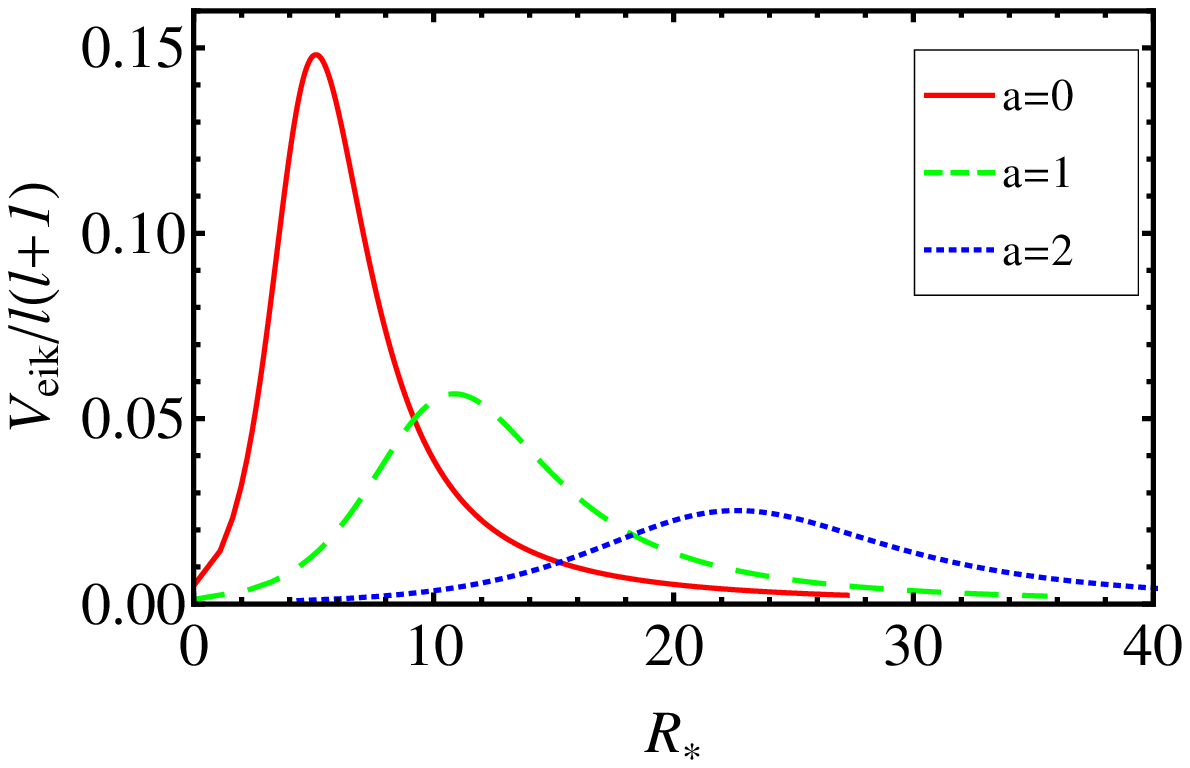}
\end{tabular}
\caption{Color online. The eikonal part of the effective potential $V_{eik}=\mathcal{V}$ for a dyon black hole solution. Left panel: The 
reduced (eikonal)
effective potential $\mathcal{V}/l(l+1)$ as a function of $R$. Right panel: The reduced (eikonal) effective potential $\mathcal{V}/l(l+1)$ as a function of $R_*$. In both panels the solid red, dashed green and dotted blue curves correspond to $a=0$, $a=1$ and $a=2$ cases, respectively. For numerical goals  here we adopted $P=1$ and $\mu=1/2$. }
\label{fig:effpot}
\end{figure*}

In this section we derive quasinormal modes (in eikonal approximation) for our static 
and spherically symmetric solution with the metric given (initially) in the following general form
\begin{equation}\label{QNM2m}
 ds^2 =-A(u) dt ^2 + B(u) du ^2 + C(u) d \Omega ^2 \ ,
\end{equation}
where $A(u),$ $B(u),$ $C(u)>0$  and $d \Omega^2 = d \theta^2 + \sin^2 \theta d \phi^2$. 
Note that in this section and below we  use the Planck units, i.e. we put 
$\hbar = G=c=1$.

We consider  a test massless scalar field defined in the background given
by the metric \eqref{QNM2m}. The equation of motion in general is written in the form of the 
covariant Klein-Fock-Gordon equation
\begin{equation}\label{QNM2}
\Delta \Psi =\frac{1}{\sqrt{\vert g \vert }}\partial_{\mu 
}(\sqrt{\vert g\vert }g^{\mu \nu } \partial_{\mu } \Psi )=0.
\end{equation}
where $\mu, \nu=0, 1, 2, 3$.
In order to solve this equation we separate variables in function $\Psi$ as follows
\begin{equation}\label{QNM3}
\Psi = e^{-i \omega t}e^{-\gamma}\Psi_{*}(u)Y_{lm},
\end{equation}
where $Y_{lm}$ are the spherical harmonics. Equation \eqref{QNM2}, after using \eqref{QNM3} yields
the equation describing the radial function $\Psi_{*}(u)$ and having a Schr\"odinger-like form
\begin{equation}\label{}
\frac{d^2 \Psi_*(u)}{du^2}+\bigg\{\frac{B}{A}\omega^{2}-\frac{B}{C}l(l+1)-\gamma''-(\gamma ')^{2}\bigg\} \Psi_*(u)=0
\end{equation}
where
\begin{equation}
\gamma =\frac{1}{2}\ln (B^{-1}C\sqrt{AB})
\end{equation}
and 
$\gamma'=d\gamma/du$, and $l$ is the multipole quantum number, $l = 0,1, \dots$. 

Taking into account above expressions one can examine a dyon-like black hole solution 
which has the following form
\begin{equation}
 ds^2 =-f(R) dt ^2 + \frac{dR ^2}{f(R)} + H^{a} R^2 d \Omega ^2 \ , \\
\end{equation}
where $f(R)$ and $C(R)$ according to Eq.\eqref{i4.10} can be written as
\begin{eqnarray}
 f(R)&=&A=H^{-a}\left( 1-\frac{2\mu}{R} \right)\ , \\
 C&=&H^{a} R^2\ ,
\end{eqnarray}
where  $H(R)= 1 + P/R$
is the moduli function, $\mu$, $P > 0$ and $ 0<a \leq 2$ as shown earlier. 
After using the ``tortoise'' coordinate transformation
\begin{equation}
dR_*=\frac{dR}{f}
\end{equation}
the metric takes the following form
\begin{equation}
 ds^2 =-f dt ^2 + f dR_*^2 + C d \Omega ^2 \ . \\
\end{equation}
For the choice   of the tortoise coordinate as a radial one ($u =R_*$) we have $A=B=f$ and
\begin{equation}
\gamma =\frac{1}{2}\ln C=\frac{1}{2}\ln (H^{a} R^2).
\end{equation}
Thus, the Klein-Fock-Gordon equation becomes
\begin{equation}\label{QNM5}
\frac{d^{2}\Psi_{*}}{dR_*^{2}}+\big\{ \omega 
^{2}-V\big\} \Psi_{*}=0, 
\end{equation}
where $\omega$ is the (cyclic) frequency of the quasinormal mode and 
$V =V(R) = V(R(R_*))$ is the effective potential
\begin{eqnarray}\label{QNM5V}
V&=&\mathcal{V}+\delta\mathcal{V},
 \\ \mathcal{V} &=&\frac{l(l+1)f}{H^{a}R^2},
 \\ \delta\mathcal{V} &=& \gamma''+(\gamma ')^{2} = \frac{ f^{2}a(a-2)P^2}{4H^{2}R^4} \nonumber
 \\&+&\frac{f (2 R-(a-2) P) (a P (R-2 \mu )+2 \mu  (P+R))}{2 H^{a+2} R^5},
\end{eqnarray}
so that $\mathcal{V}$ is the eikonal part of the effective potential. 
Here and below we denote $F' = \frac{dF}{dR_{*}}= f \frac{dF}{dR}$.

In what follows we consider the so-called eikonal approximation 
 when $l \gg 1$. 

The maximum of the eikonal part of the effective potential is found from the condition
\begin{eqnarray} \label{QNM6V}
\mathcal{V}'&=& f \frac{d \mathcal{V}}{dR} \quad  \nonumber \\
&=& - 2 f [R^2 + ((1-a)P - 3 \mu) R + (2a - 3) P \mu] \times \nonumber \\
 &&\times (H^{-2a -1}R^{-5}) = 0, \quad 
\end{eqnarray}
or, equivalently,
\begin{equation} \label{QNM7R}
 R^2 + ((1-a)P - 3 \mu) R + (2a - 3) P \mu =0,
\end{equation}
which yields the corresponding radius
\begin{eqnarray}
R_0&=&\frac{a-1}{2}P+\frac{3\mu}{2}+\frac{1}{2}\sqrt{\mathcal{D}}, \label{QNM8R}  \\
\mathcal{D}& =& (1-a)^2 P^2+2(3-a)P\mu+9\mu^2 > 0. \label{QNM8D}
\end{eqnarray}
The inequality (\ref{QNM8D}), which is valid for  $0 < a \leq 2$,
is a trivial one.

It may be readily verified by using quadratic equation for 
$Z = R - 2 \mu$ 
\begin{equation}  \label{QNM7z}
Z^2 + ((1-a) P + \mu) Z - 2 \mu^2   - P \mu = 0
\end{equation}
and $\mathcal{D} > 0$ that 
\begin{equation} \label{QNM7zR0}
  R_0 = R_{0,+} > 2 \mu > R_{0,-},
\end{equation}
for  all $0 < a \leq 2$. Here $R_{0,-}$ is
another root  of the quadratic equation (\ref{QNM7R}) , 
which corresponds to the ``location'' under the horizon and 
 is irrelevant for our consideration. 

\begin{figure*}[ht]
\centering
\begin{tabular}{lr}
\includegraphics[width=0.48\linewidth]{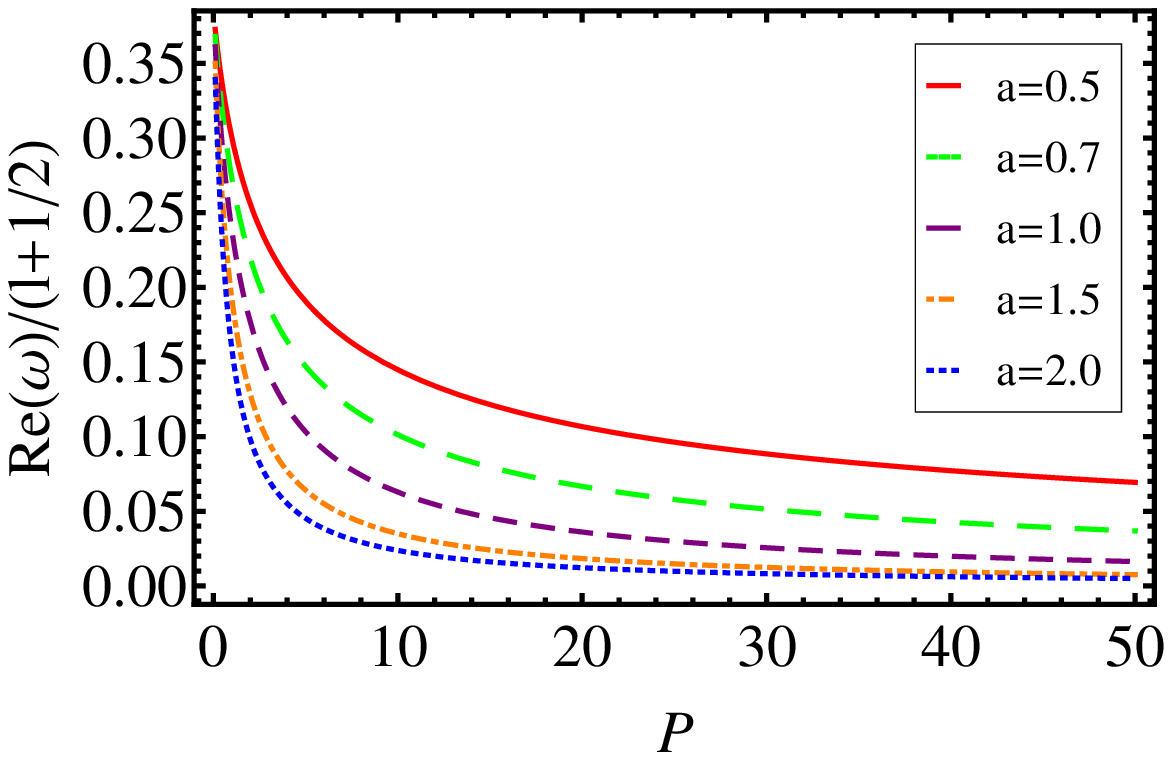} &
\includegraphics[width=0.48\linewidth]{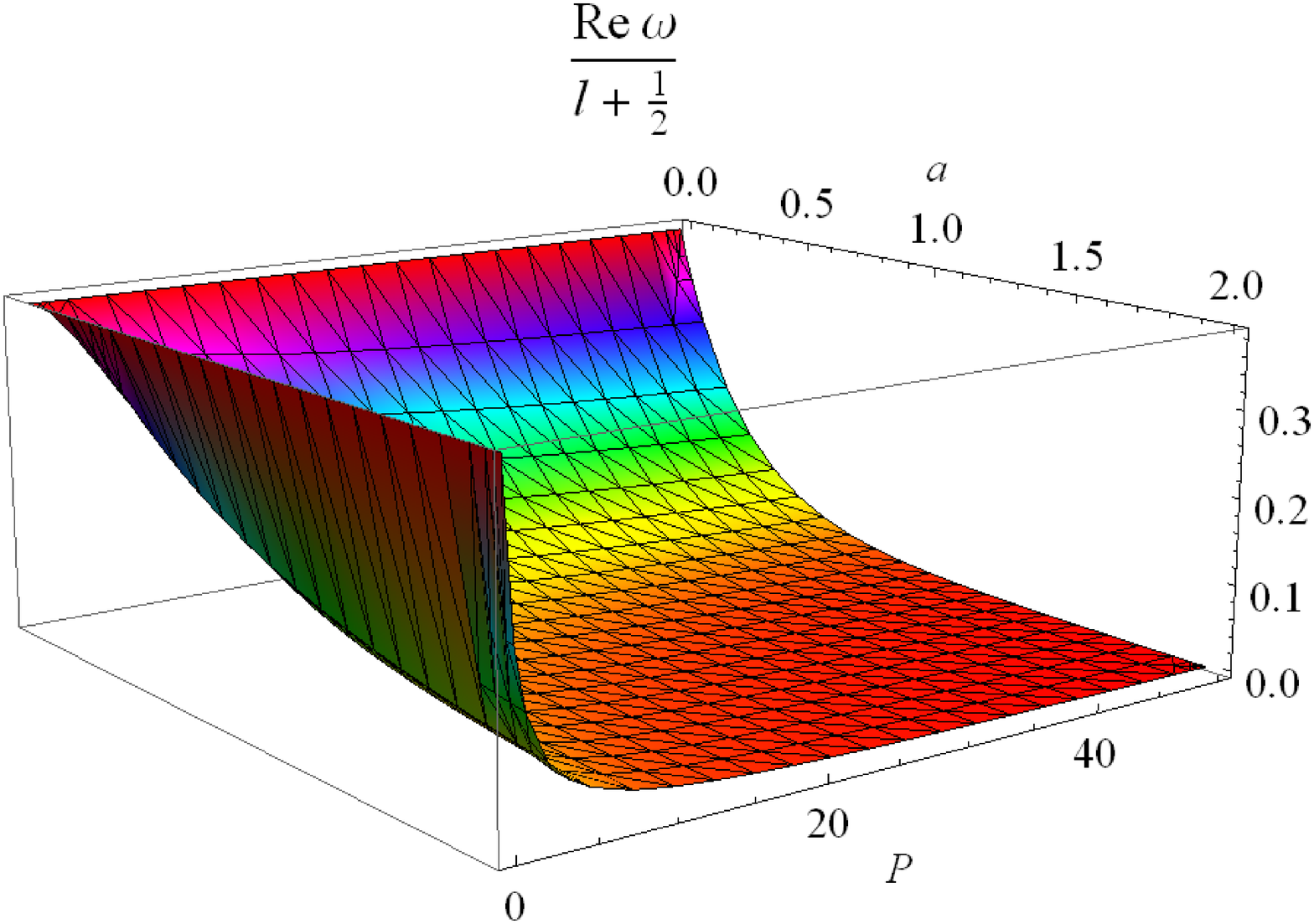}
\end{tabular}
\caption{Color online. The dependence of $Re(\omega)$ on $P$ for different $a$ in the range $0 <a\leq 2$. Here we adopt $\mu=1/2$.
Left panel: two dimensional plot. Right panel: three dimensional plot. }
\label{fig:ReOmega}
\end{figure*}

\begin{figure*}[ht]
\centering
\begin{tabular}{lr}
\includegraphics[width=0.48\linewidth]{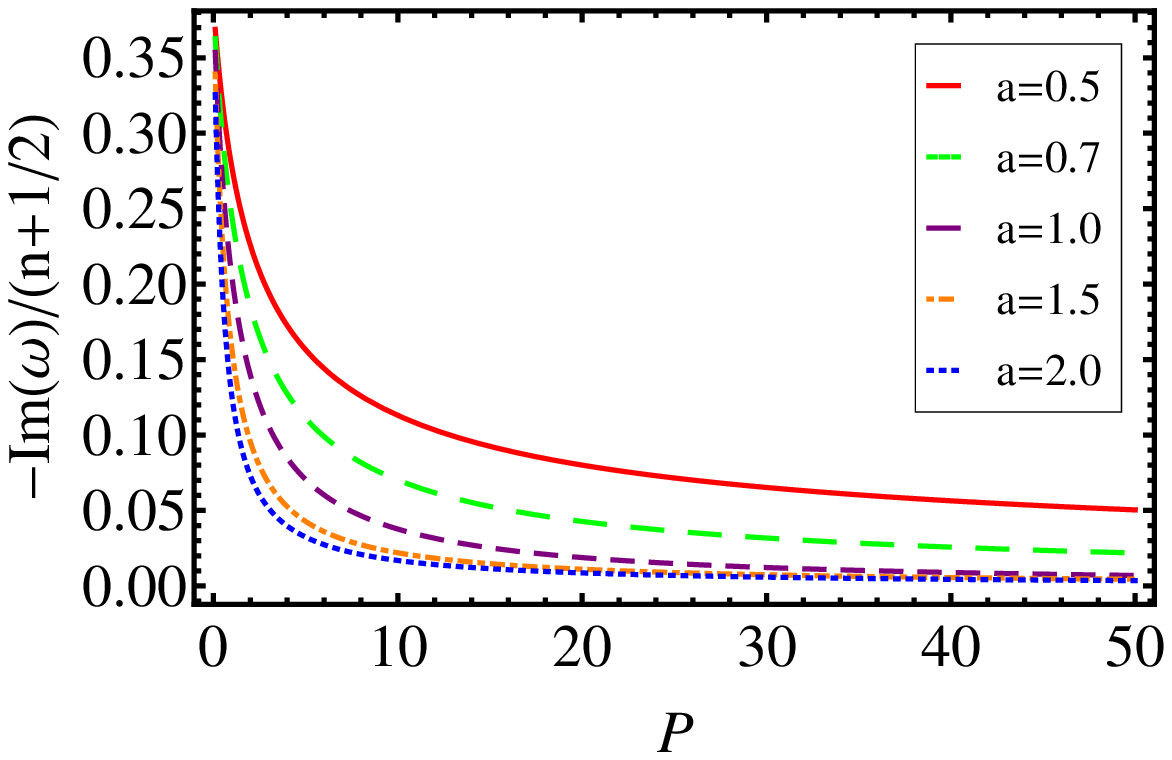} &
\includegraphics[width=0.48\linewidth]{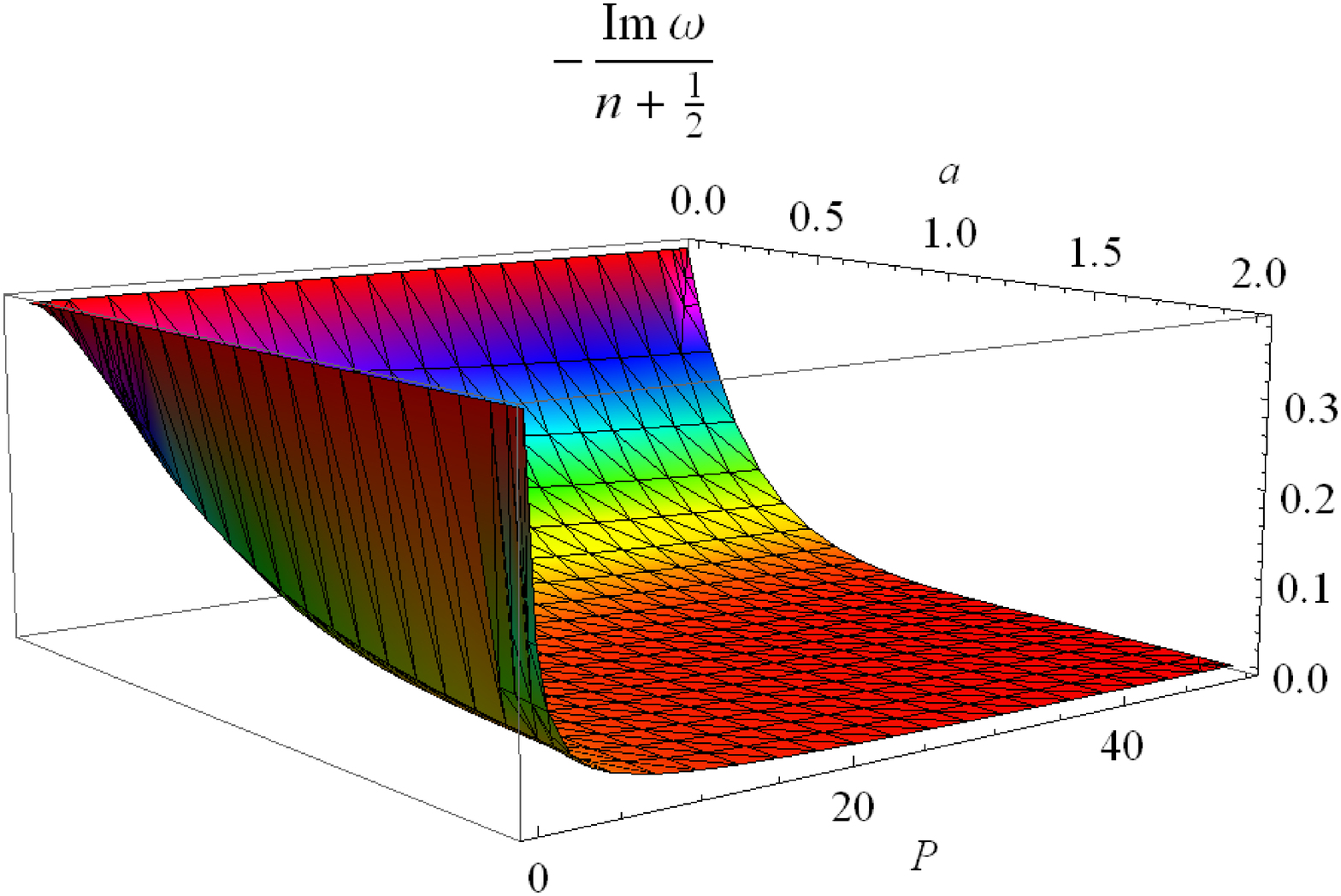}
\end{tabular}
\caption{Color online. The dependence of $-Im(\omega)$ on $P$ for different $a$ in the range $0 <a\leq 2$. Here we adopt $\mu=1/2$.
Left panel: two dimensional plot. Right panel: three dimensional plot. }
\label{fig:ImOmega}
\end{figure*}

The maximum of the eikonal part  of the effective potential thus becomes
\begin{equation}
\mathcal{V}_0=\mathcal{V}(R_0)=
\frac{l (l+1) }{R_0^2}\left(1-\frac{2 \mu }{R_0}\right) \left(1+\frac{P}{R_0}\right)^{-2 a}. \label{QNM9V}
\end{equation}

In Fig.~\ref{fig:effpot} we plot the reduced eikonal part of the effective potential $\mathcal{V}/(l(l+1))$ ($l \neq 0$) as a function of the radial coordinate $R$ (left panel), and tortoise coordinate $R_*$ (right panel). 
As can be seen from examples presented in figure for special fixed values of $P$ and $\mu$,  the maximum of the effective potential is largest for $a=0$ case and smallest for $a=2$ case. The case with $a=1$ is in the middle. At large distances the effective potential tends to zero, as expected.

The second derivative with respect to the tortoise coordinate is given by
\begin{eqnarray}
\mathcal{V}_0''&=&\frac{d^2\mathcal{V}}{dR_*^2}\bigg|_{R_{*}=R_{*}(R_0)}=f^2 \frac{d^2\mathcal{V}}{dR^2}\bigg|_{R=R_0}
=  -\frac{2 l (l+1) }{R_0^5} \times \nonumber \\
&\times& \left(1-\frac{2\mu}{R_0}\right)^2 \left(1+\frac{P}{R_0}\right)^{-(1+4a)} \sqrt{\mathcal{D}}, \label{QNM10V2}
\end{eqnarray}
where $\mathcal{D}$ is defined in  (\ref{QNM8D}).

The square of the cyclic frequency in the eikonal approximation reads as following \cite{2009CQGra..26p3001B,2011RvMP...83..793K}
\begin{equation}
\omega^{2}=\mathcal{V}_0-i \left(n+\frac{1}{2}\right) \sqrt{-2 \mathcal{V}_0''} + O(1),
\end{equation}
where $l \gg 1$ and $l \gg n$. Here $n = 0, 1, \dots$ is the overtone number. By choosing an appropriate
sign for  $\omega$ we get the asymptotic relations (as $l \to + \infty$) 
on real and imaginary parts of complex  $\omega$ in the eikonal approximation
\begin{eqnarray}
  {\rm Re}(\omega)  &=&   \left(l + \frac{1}{2}\right) H_{0}^{-a} F_0^{1/2} R_0^{-1}
      + O\left(\frac{1}{l+\frac{1}{2}}\right),  \label{QNM11Re} 
      \\
 {\rm Im}(\omega)  &=&  - \left(n + \frac{1}{2}\right) H_{0}^{-a - 1/2} F_0^{1/2} R_0^{-3/2} \mathcal{D}^{1/4} \nonumber \\
    &+& O\left(\frac{1}{l+\frac{1}{2}}\right), \label{QNM11Im}
\end{eqnarray}
where $H_0 = 1+\frac{P}{R_0}$, $F_0= 1-\frac{2\mu}{R_0}$ and $R_0, \mathcal{D}$ are given by (\ref{QNM8R}), (\ref{QNM8D}), respectively.

In Fig.~\ref{fig:ReOmega} we constructed real part of the QNM frequency, Re$(\omega)$, as a function of parameter $P$ for different values of $a$ and $\mu=1/2$. 
In the right panel we have a three-dimensional plot of Re$(\omega)$ versus $P$ and $a$. Here one can notice that in the limiting case when $a=0$ we recover constant Re$(\omega)$ identical to the case of the \\ 
Schwarzschild black hole.}

In Fig.~\ref{fig:ImOmega}
we constructed the imaginary part of the QNM frequency with negative sign, -Im$(\omega)$, as a function of parameter $P$ for different values of $a$ and $\mu=1/2$ in analogy to Fig.~\ref{fig:ReOmega}. At first sight Figs. ~\ref{fig:ReOmega} and \ref{fig:ImOmega} seem similar. However, according to Eqs.~\eqref{QNM11Re} and \eqref{QNM11Im} this is not the case.

{\bf Remark.} {\it It was shown in Ref. \cite{2009PhRvD..79f4016C}  that parameters of the unstable circular null geodesics around  stationary spherically symmetric and asymptotically flat black holes are in correspondence with the eikonal part of  quasinormal modes of these black holes.
See also \cite{2000PhRvD..62h4003V,2016PhRvD..94j6005C} and references therein. 
But as it was pointed out in Ref. \cite{2017PhLB..771..597K} this 
correspondence is valid if: (a) perturbations are described by a ``good'' effective potential,  ( b) ``one is limited by perturbations of test fields only, and not of the gravitational field itself or other fields, which are non-minimally coupled to gravity.''  Here we do not consider this correspondence for our solution, postponing  this  to future publication.}

\section{Limiting cases corresponding to the Schwarzschild and Reissner-Nordstr\"om black holes}\label{limitingcase2}

In this section we consider two  limiting cases $a=+0$ and $a=2$ corresponding to the Schwarzschild and  \\ Reissner-Nordstr\"om    metrics, respectively.

a) Let us first consider the case when $a = +0$. This limit may be obtained 
   in a strong coupling regime when
   \begin{equation} \label{QNM8L0}
   \vec{\lambda}_1 = \lambda \vec{e_1}, \qquad \vec{\lambda}_2 = \lambda \vec{e_2},
 \end{equation}
where $\vec{e_1}^2 = \vec{e_1}^2 = 1$, and 
 \begin{equation} \label{QNM8LS}
   \vec{e_1}\vec{e_2} \neq \pm 1, \quad \lambda \to + \infty.
 \end{equation} 

In this case  the relations (\ref{QNM11Re}) and (\ref{QNM11Im}) for QNM in the eikonal approximation read as follows
\begin{eqnarray}
 {\rm Re}(\omega)&=&\left(l+\frac{1}{2}\right)\sqrt{\frac{M}{r_0^3}}+O\left(\frac{1}{l+\frac{1}{2}}\right),\\
{\rm Im}(\omega)&=&-\left(n+\frac{1}{2}\right)\sqrt{\frac{M}{r_0^3}}+O\left(\frac{1}{l+\frac{1}{2}}\right),
\end{eqnarray}
where $r_0 = R_0 = 3M$ corresponds  the position where the black-hole effective potential attains its maximum. Note that $r_0=3M$  is the radius of the photon sphere 
for  the  Schwarzschild black-hole. These results have been obtained in Ref.~ \cite{1984PhLA..100..231B} and our outcomes are consistent with them.

b) Now let us consider the case when $a = 2$. As was mentioned above this takes
   place for collinear vectors $\vec{\lambda}_1$, $\vec{\lambda}_2$. 
   One can also obtain the limit $a = 2$ in the weak coupling regime when dilatonic coupling vectors obey (\ref{QNM8L0}) and 
    \begin{equation} \label{QNM8LW}
   \vec{e_1}\vec{e_2} \neq - 1, \quad \lambda \to + 0.
 \end{equation}  

In this case the eikonal QNM (see (\ref{QNM11Re}) and (\ref{QNM11Im})) read 

\begin{eqnarray}
{\rm Re}(\omega)&=&\left(l+\frac{1}{2}\right)\sqrt{\frac{M}{r_0^3}-\frac{Q^2}{2 r_0^4}}+O\left(\frac{1}{l+\frac{1}{2}}\right),\\
{\rm Im}(\omega)&=&-\left(n+\frac{1}{2}\right)\sqrt{\frac{M}{r_0^3}-\frac{Q^2}{2 r_0^4}} \sqrt{\frac{3 M}{r_0}-\frac{2 Q^2}{r_0^2}}\nonumber
\\&+&O\left(\frac{1}{l+\frac{1}{2}}\right), 
\end{eqnarray}
where $r_0=3M/2+ (1/2)\sqrt{9M^2-4Q^2} = R_0 + P$ corresponds to the position of the unstable, 
circular photon orbit in the Reissner-Nordstr\"om spacetime. 
These results have been obtained in Ref.~ \cite{1996PhRvD..54.7470A} (for $n =0$) and our outcomes are compatible with them when the relation (for our notation) $Q^2=2Q_{RN}^2$ is applied.

\section{Hod conjecture}\label{Hod}

Here we test/check  the conjecture formulated by Hod \cite{Hod} on the existence of quasi-normal modes obeying the inequality 
\begin{equation} \label{QNM13H}
  |{\rm Im}(\omega)  | \leq \pi T_H,
\end{equation}
where $T_H$ is the Hawking temperature.

Recently the Hod conjecture has been tested in theories with higher curvature corrections such as the 
\newline
Einstein–Dilaton–Gauss–Bonnet and Einstein–Weyl for the Dirac field \cite{Zinhailo2019}. It has been shown that in both theories the Dirac field obeys the Hod conjecture for the whole range of black-hole parameters \cite{Zinhailo2019}.

Here we test/check this conjecture  by using eikonal relations  (\ref{QNM11Im}) 
for  ${\rm Im}(\omega) $
and the relation for the Hawking temperature (\ref{i5.2}). For our purpose it is sufficient
to check the validity of the inequality 
\begin{eqnarray} 
 y \equiv \frac{|{\rm Im}(\omega_{eik})(n=0) |}{\pi T_H} = 4 \left(1 + \frac{p}{x}\right)^{ - a - 1/2} \times  \nonumber \\ 
 \times  \left(1 - \frac{2}{x}\right)^{ 1/2} x^{-3/2} d^{1/4} \left(1 + \frac{p}{2}\right)^{ a} < 1, \label{QNM14y}
\end{eqnarray}
for all $p = P/\mu > 0$, where 
\begin{eqnarray}
x & \equiv & R_0/\mu = \frac{a-1}{2} p+\frac{3}{2}+\frac{1}{2}\sqrt{d}, \label{QNM15x}  \\
d  & \equiv & \mathcal{D}/\mu^2 =   (1-a)^2 p^2+ 2(3-a)p +9 > 0.   \label{QNM15d}   
\end{eqnarray}
In (\ref{QNM14y}) we use the limiting  ``eikonal value'' given by the first term in (\ref{QNM11Im}) for
the lowest overtone number $n=0$. 

In Table~\ref{table1} we present the results for the numerical  testing of the Hod bound  by using obtained relations for the eikonal QNM in the ground state $n$=0. It turned out that the Hod bound is valid (in the eikonal regime)  in the range $0<a \leq1$. There are maximum $y_{max} = y_{max}(a)$ and limiting 
$y_{lim}= y_{lim}(a)$ values of function $y(p,a)$ for different values of parameter $a$ in the considered range. 

\begin{table}[ht]
\begin{center}
\caption{Maximum and limiting values of $y$ for various values of the model parameter $a$ in the ground state $n$=0. 
$y_{max}$ is the maximum of $y$, $p_0$ is the value of $p$ corresponding to $y_{max}$ and $y_{lim}=\lim\limits_{p\to \infty}y$ .  }
\vspace{3 mm}
\label{table1}
\begin{tabular}{cccc}
\hline
$a$ & $p_0$   & $y_{max}$    & $y_{lim}$   \\
\hline

10$^{-2}$ & 1.314 & 0.770 & 0.769 \\

0.1   & 1.421 & 0.772 & 0.762  \\

0.2   & 1.563 &	0.774 &	0.752  \\

0.4   & 1.954 &	0.780 &	0.726  \\

0.5   & 2.233 & 0.785 & 0.707  \\

0.6   & 2.608 &	0.790 &	0.681  \\

0.8   & 3.933 &	0.808 &	0.588  \\

1.0   & 8.196 & 0.847 & 0     \\

  \hline
    \end{tabular}
  \end{center}
\end{table}

\begin{figure*}[ht]
\centering
\begin{tabular}{lr}
\includegraphics[width=0.48\linewidth]{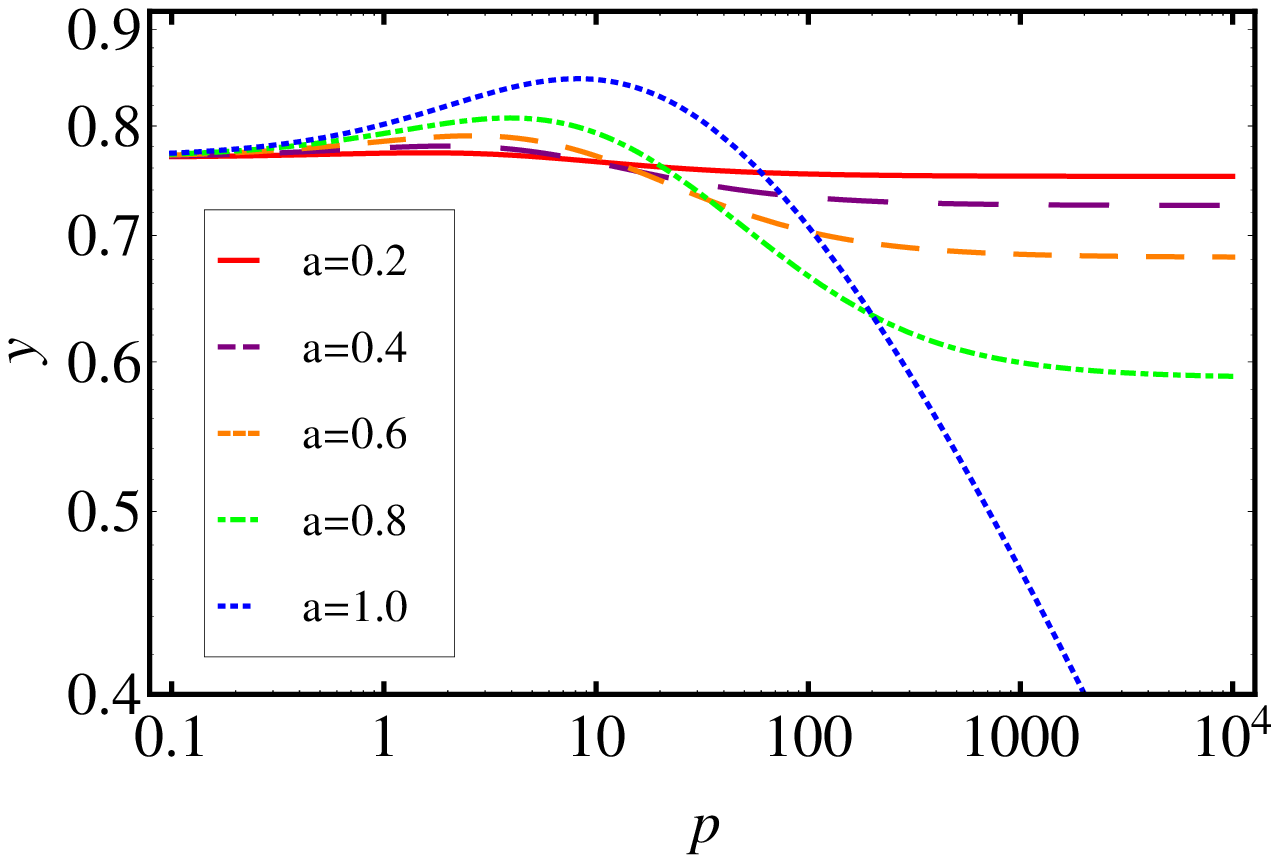} &
\includegraphics[width=0.48\linewidth]{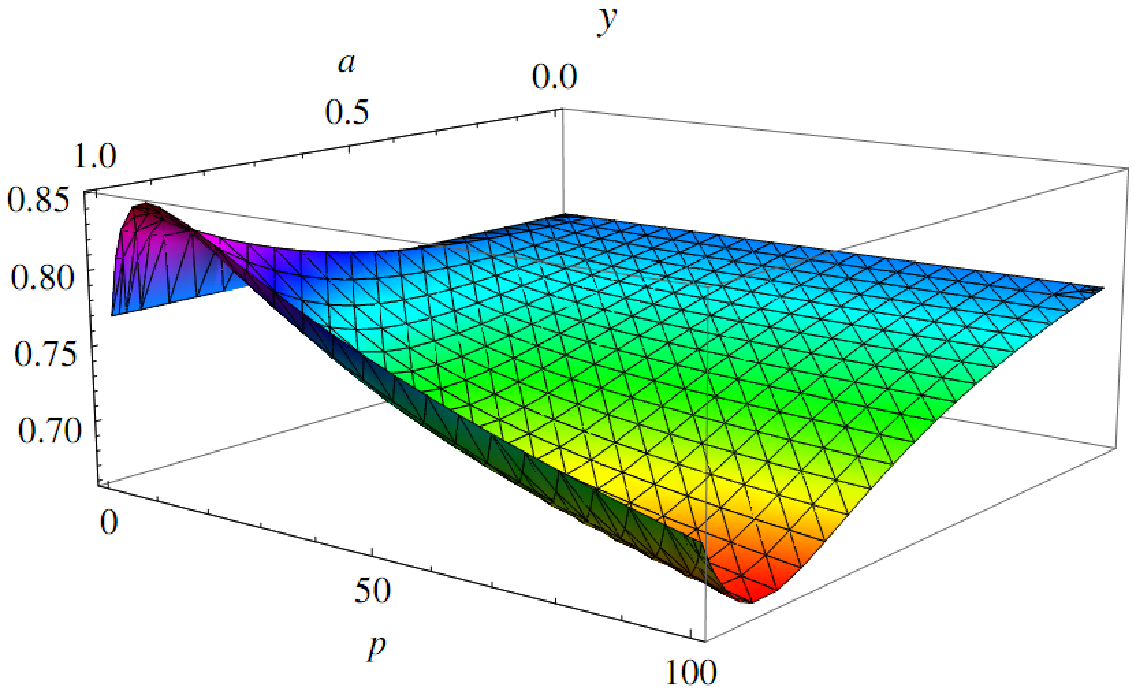}
\end{tabular}
\caption{Color online. The dependence of $y$  on $p$ for different $a$ (see (\ref{QNM14y})) 
in the range $0<a \leq 1$. Left panel: 
two dimensional plot in logarithmic scale. Right panel: three dimensional plot. }
\label{fig:ypPlots}
\end{figure*}

It may be verified that
\begin{equation} \label{QNM16ylim}
 y_{lim}(a) = \left(\frac{1 - a}{3 - 2a }\right)^{3/2 -a} 2^{2-a} <  \frac{4}{3\sqrt{3}} <1
 \end{equation}
for $0 < a < 1$ and $y_{lim}(1) =0$. The relation for $y_{lim}(a)$
  just follows from 
 \begin{equation} \label{QNM16ylim2}
 \lim_{p \to + \infty} x(p,a) = (3 - 2a)/(1-a) 
 \end{equation}
for $0 < a < 1$.

We denote the value of $p$ corresponding to $y_{max}(a)$ as $p_0 = p_0(a)$. For increasing $a$, 
the values of $p_0(a)$ and $y_{max}(a)$ increase and $y_{lim}(a)$ decrease. We obtain $y_{max}(1) \approx 0.847$ and $y_{lim}(+0) =0$. 
For decreasing $a$, both $y_{max}(a)$ and $y_{lim}(a)$ approach a finite value, corresponding to the Schwarzschild case $\frac{4}{3\sqrt{3}} \approx 0.7698$ when $a \to 0$.

In Fig.~\ref{fig:ypPlots} we illustrate $y = y(p,a)$ as a function of $p$. In the left panel we build two dimensional plot 
for $a$ = 0.2, 0.4, 0.6, 0.8, 1.0 and in the right panel we construct three dimensional plot for the range 
$0 < a \leq 1$, 
where the Hod conjecture holds. Thus, we are led to the following proposition.

{\bf Proposition.} The dimensionless parameter $y = y(p,a)$ from (\ref{QNM14y})  obeys the inequality: 
$y(p,a) < 1$ for all $p > 0$ and $a \in (0,1]$.

For $0 < a < 1$ this proposition is proved analytically in Appendix. For $a=1$ it is justified by our numerical bound $y < y_{max}(1) \approx 0.847$.

\begin{figure*}[ht]
\centering
\begin{tabular}{lr}
\includegraphics[width=0.48\linewidth]{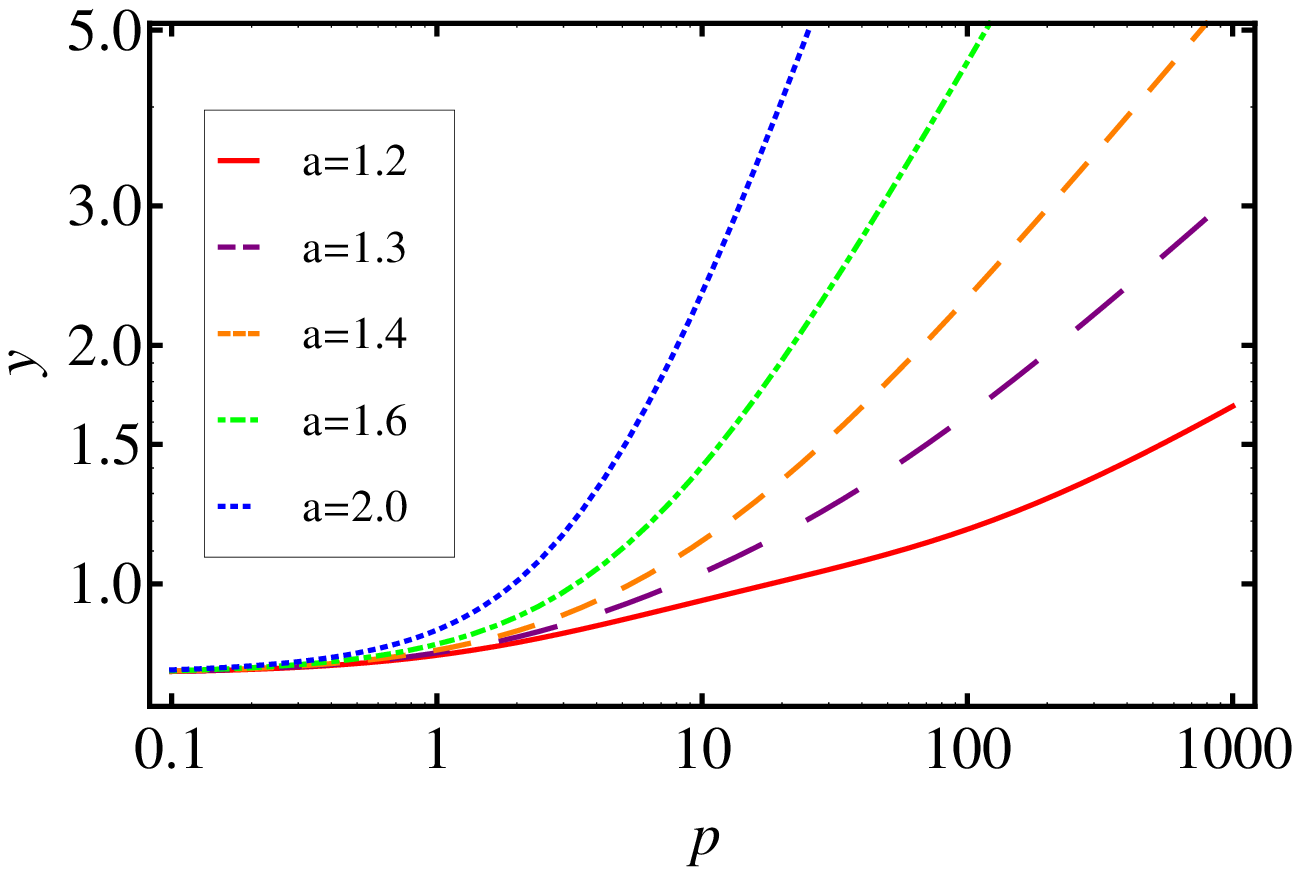} &
\includegraphics[width=0.48\linewidth]{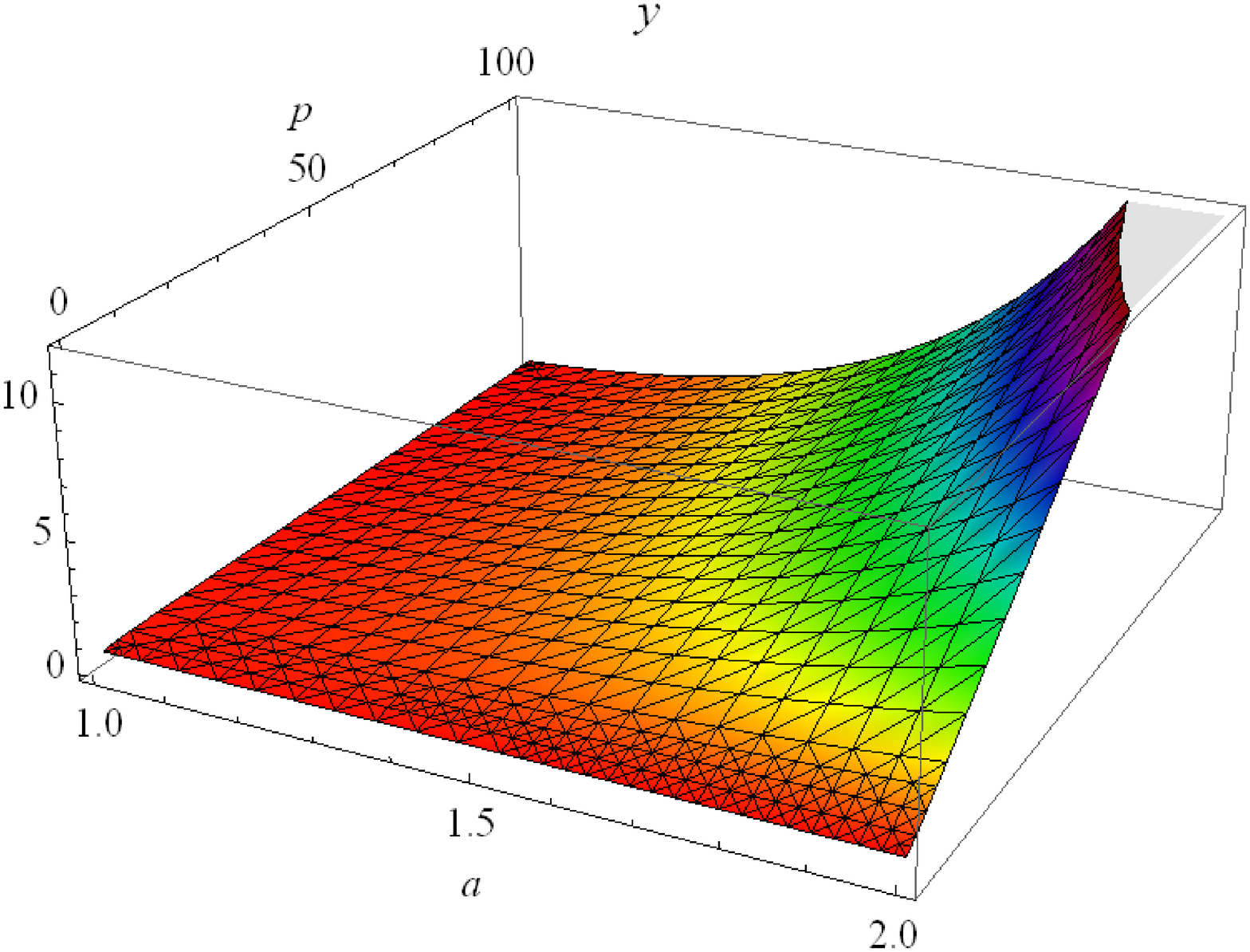}
\end{tabular}
\caption{Color online. The dependence of $y$ on $p$ for different $a$ (see (\ref{QNM14y})  
in the range $1<a\leq 2$.
Left panel: two dimensional plot in logarithmic scale. Right panel: three dimensional plot. }
\label{fig:ypPlotsa2}
\end{figure*}

In addition we considered the range $1< a \leq 2$ for testing the Hod conjecture. 
In this case we have 
\begin{equation} \label{QNM17ylim}
 y_{lim}(a) = \lim_{p \to + \infty} y(p,a) = + \infty  
\end{equation}
due to asymptotic relation  $y(p,a) \sim C(a) p^{a-1}$ for $p \to  + \infty $ following from $x(p,a) \sim (a-1) p$, where $C(a) = 2^{2-a} (a/(a-1))^{-a - 1/2} (a-1)^{-1}$. Strictly speaking for $1 < a \leq 2$ there exist critical values $p_{crit}(a)$ of parameter $p$ such that for $p \in (0, p_{crit}(a))$ the  Hod conjecture holds in eikonal regime while for $p \in (p_{crit}(a), + \infty)$ it fails (see Fig.~\ref{fig:ypPlotsa2} for details). Here the limit $p = + \infty$ corresponds to extremal (black hole) case which is not considered here.

{\bf Remark.} {\it Recently, in Ref. \cite{2016PhRvD..93j4053C} some example of the violation of the Hod conjecture has been found for certain (scalar gravitational) perturbations around $D = 5$ 
Gauss-Bonnet-de Sitter black hole  solution.}

In Table~\ref{table2} we present  some
critical values   $p_{crit}$, which are obtained through the condition $y(p_{crit})=1$ 
(with $y$ calculated for  the ground state $n$=0)
and $q_{crit}$ corresponding to $p_{crit}$ according to Eq.~\eqref{eq:qcrit} for various values 
of the model parameter $a$ obeying $1< a \leq 2$.

Here we use the following relation (see Eq.~\eqref{i5.1fpq})
\begin{equation}\label{eq:qcrit}
    q=\frac{|Q|}{M}=\frac{2\sqrt{2}\sqrt{p(2+p)}}{2+ap} < \frac{2\sqrt{2}}{a} = q_{ext},
\end{equation}
where $q_{ext}$ corresponds to extremal case. (Here $G = 1$.)

\begin{table}[ht]
\begin{center}
\caption{Critical values of $p$ denoted as $p_{crit}$, obeying $y(p_{crit})=1$, 
 and $q_{crit}$  corresponding  to $p_{crit}$ according to Eq.~\eqref{eq:qcrit} 
 for certain values of  $a$ ($1<a \leq 2$). }
\vspace{3 mm}
\label{table2}
\begin{tabular}{ccc}
\hline
$a$ & $p_{crit}$   & $q_{crit}$   \\
\hline

1.1 & 8131.908 & 2.571  \\

1.2   & 18.157 & 2.274 \\

1.3   & 8.031 &	2.041   \\

1.4   & 5.402 &	1.870   \\

1.6   & 3.356 & 1.627   \\

1.8   & 2.461 &	1.457  \\

2.0   & 1.951 &	1.330  \\

  \hline
    \end{tabular}
  \end{center}
\end{table}

In Fig.~\ref{fig:ypPlotsa2} we illustrate $y = y(p,a)$ as a function of $p$. In the left panel we build two dimensional plot for $a$ = 1.2, 1.3, 1.4, 1.6, 2.0 and in the right panel we construct three dimensional plot for the range $1 < a \leq2$. In this case the Hod 
inequality (\ref{QNM14y}) holds  in the range  $p \in (0, p_{crit}(a))$,  while for $p \in (p_{crit}(a), + \infty)$ it breaks.

{\bf Remark.} {\it  In Ref.~\cite{2012PhLB..710..349H}, the eikonal QNM  frequencies for charged scalar field  in the space-time of a charged Reissner-Nordstr\"om black hole were obtained analytically in the regime $l^2 \geq Q q_{*} \geq l$, where $Q$ is the electric charge of the black hole and $q_{*}$ is the electric charge of the field. In this regime the obtained fundamental frequencies were shown to saturate the Hod bound. It should be noted that this result can not be applied to our analysis for $a=2$, since we deal here with $q_{*} = 0$ case.}

\section{Conclusions}\label{conclusion}

We have examined a non-extremal black hole dyon-like solution in a 4-dimensional gravitational model with two scalar fields and two Abelian vector fields proposed in Ref.~\cite{2020JPhCS1690a2143B}. The model contains two vectors of dilatonic coupling vectors $\vec{\lambda}_s \neq \vec{0}$, $s =1,2$, obeying $\vec{\lambda}_1 \neq - \vec{\lambda}_2$ and additional relations   (\ref{i4.8bdd}). We have also presented some physical parameters of the solutions: gravitational mass $M$, scalar charges $Q_{\varphi}^i$, Hawking temperature, black hole area entropy. In addition, we considered the first law of black hole thermodynamics and checked the validity of the Smarr relation for our model.

In fact this is a special solution with dependent electric and magnetic charges, see (\ref{i4.8c}). In the case of non-collinear vectors $\vec{\lambda}_1$,  $\vec{\lambda}_2$ the metric of the solution describes a black hole with one (external horizon) and singularity hidden by it. For  collinear vectors $\vec{\lambda}_1$,  $\vec{\lambda}_2$ the metric coincides with the  Reissner-Nordstr\"om   metric possessing two horizons and hidden singularity. 

We  have studied the solutions  to  massless (covariant)  Klein-Fock-Gordon equation in the background of our static and spherically symmetric metric, by using the variable separation method. The Klein-Fock-Gordon equation is simplified in the  tortoise coordinate leading to  radial equation governed by  effective potential. This potential contains the parameters of solution such as $P >0$, $\mu >0$ and dimensionless parameter $a \in (0,2]$  depending upon the coupling vectors $\vec{\lambda}_s$ which are the initial parameters of the model. The physical quantities, such as  mass, color charges and scalar charges  contain some  of these  parameters. 
 
Here we mainly focused on the eikonal part of the effective potential and calculated the value of the radial coordinate (radius) $R_0$ corresponding to the maximum of this part of the effective potential. Knowing the maximum of the eikonal part of the effective potential and corresponding radius, we have calculated the cyclic frequencies of the quasinormal modes  in the eikonal approximation. We have also  considered two limiting cases reducing to the Schwarzschild and Reissner-Nordstr\"om solutions, when the  parameter of the solution $a$ accepts two distinct values, i.e. $a=+0$ and $a=2$, respectively. Thus, we  have made sure that our outcomes are consistent with the previous results in the literature.

We have also tested the validity of the Hod conjecture for  our solution by using QNM  frequences in the eikonal approximation with the lowest value of the overtone number $n=0$. 
It turned out that the Hod assumption holds  in the range of $0<a\leq1$.  The conjecture is valid in this range  since it is supported by  examples of states with large enough values of angular number $l$.
For $1 < a \leq 2$ we have found that the Hod bound is satisfied for $n=0$, and small enough values of charge $Q$ ($Q/M < q_{crit}(a)$) and big enough values of $l$ ($l \gg 1$). 

It would be interesting to explore in detail  QNM frequencies  in the vicinity of  $a=0$ and $a=2$,  by using the treatment of Ref.~\cite{2019EPJC...79..629C}, e.g. extending the results for the \\ Schwarzschild and Reissner-Nordstr\"om  solutions by using expansion in a small parameter 
($a$ or $a-2$). Another issue of interest is  the numerical calculation of QNM frequencies by using higher-order WKB formula for certain lower levels (labelled by $l$ and $n$), see Ref.~\cite{2019CQGra..36o5002K}, e.g. verifying the Hod conjecture for $1< a < 2$, calculating grey-body factors etc.  All these issues may be addressed in our future studies.

\begin{acknowledgements}
 This paper has been supported by the RUDN University Strategic Academic Leadership Program 
 (recipient: V.D.I. - mathematical model development) and by the Ministry of Education and 
 Science of the Republic of Kazakhstan, project identification registration number (IRN):
 AP08052311 (recipients K.B. - simulation model development and numerical checking of results). The authors thank Roman Konoplya for fruitful discussions and valuable comments during the preparation of the manuscript.
\end{acknowledgements}

\appendix
\section{Analytic proof of the Hod conjecture }

Here we prove the Proposition from Section \ref{Hod} for $0 < a < 1$.
The quadratic equation for $x = x(p,a)$ from (\ref{QNM15x}) in terms of $a$, $p$ reads 
\begin{equation}
x^2-[(a-1) p+3)] x+(2 a-3) p=0, 
\end{equation}
see \eqref{QNM7R}. Due to  (\ref{QNM7zR0}) we have $x > 2$.
Substituting $x= u+2$,  we get 
\begin{equation}\label{Append_1}
u^2+[(1-a)p+1] u =2+p=1+\frac{t-a}{1-a},
\end{equation}
where $t=1+(1-a) p$, $t > 1$ and $u > 0$ (since $u = u_{+}$ 
is the large root of the quadratic equation \eqref{Append_1} 
and the small root $u = u_{-}$ obeys $u_{+} u_{-} = -2 -p <0$). 
See also Eq. \eqref{QNM7z} for $Z = u \mu$ ($\mu > 0$).
    
Thus, 
\begin{equation} \label{Append_20}
p/2+1= u(u+t)/2, 
\end{equation}
 see \eqref{Append_1}. 
Next, from the Eq. \eqref{Append_1}
we get 
\begin{equation} \label{Append_2}
[1-(1-a) u] p= (u-1) (u+2),
\end{equation}
which implies (due to $p > 0$, $u >0$) 
\begin{equation} \label{Append_222}
1 <  u < (1-a)^{-1},
\end{equation}
and hence
\begin{equation}
p=\frac{(u-1) (u+2)}{1-(1-a) u},
\end{equation}
so
\begin{equation}
\frac{p}{x}+1=\frac{a u}{1-(1-a) u}.
\end{equation}
Finally, from the  equation for $x$ \eqref{QNM15x} , we get
\begin{equation}
\sqrt{d}=2 x-(a-1) p-3=2 u+t.
\end{equation}
Plugging all  of that in \eqref{QNM14y}, we see that we need to 
prove the following bound 
\begin{eqnarray} \label{Append_0}
y = 4\left[\frac{(u+t)(1-(1-a) u)}{2 a}\right]^a \left[\frac{1-(1-a) u}{a u}\right]^{1/2} \times \nonumber \\
 \times \left[\frac{u}{u+2}\right]^{1/2} \frac{(t+2 u)^{1/2}}{(u+2)^{3/2}}
 <  1.
 \end{eqnarray}

Now, from \eqref{Append_1}
\begin{equation}
(1-a) u (t+u)= 1+ t -2a,
\end{equation}
so the first bracket in \eqref{Append_0} simplifies to
\begin{eqnarray}
\left(\frac{u-1+2 a}{2 a}\right)^a=\left(1+\frac{u-1}{2 a}\right)^a \nonumber \\
 \leq 1+\frac{u-1}{2}=\frac{u+1}{2}.
\end{eqnarray}
Here we have used well-known convexity inequality $(1 + v)^a \leq 1 + a v$ 
which is valid for $v >  -1$ and $0 < a \leq 1$  (for our case  $v >  -1$ 
is valid due to \eqref{Append_222}).
We can cancel $u^{1/2}$ and combine all $(u+2)$'s, which leaves us with the factor
\begin{equation}
F(t)=\frac{(1-(1-a) u) (2 u+t)}{a}.
\end{equation}
Here we use the  differentiation tool
\begin{equation}
\frac{\text{d}}{\text{dt}} \ln F (t)=-\frac{(1-a) \dot{u}}{1-(1-a) u}+\frac{2 \dot{u}+1}{2 u+t},
\end{equation}
where $\dot{u} = du/dt$. From \eqref{Append_1}, we have 
\begin{equation} \label{Append_3}
(1-a)(2 u+t) \dot{u}=1-(1-a) u, 
\end{equation}
and hence $\dot{u} > 0$ (see (\ref{Append_222})). Using \eqref{Append_3}
we obtain
\begin{equation}
\frac{\text{d}}{\text{dt}} \ln F (t)=
-\frac{1}{2u + t}+\frac{2 \dot{u}+1}{2 u+t} = \frac{2 \dot{u}}{2 u+t},
\end{equation}
so the logarithmic derivative evaluates to
\begin{equation} \label{Append_4}
\frac{\text{d}}{\text{dt}} \ln F (t) =\frac{2 \dot{u}}{2 u+t} < \frac{2 \dot{u}}{2 u+1}=\frac{\text{d}}{\text{dt}} \ln (2 u+1)
\end{equation}
(here we use $\dot{u} > 0$ and $t > 1$). 
Thus, integrating \eqref{Append_4} (in $t$ from $1$ to $t$) and using 
coincidence of initial values: $F(1) = 2 u(1) +1 =3$
($u(t) \to 1$ as $t \to 1$ and $p \to +0$) we get
\begin{equation}
 F (t) < 2 u+1
\end{equation}
and we end up with proving that
\begin{equation}
2 (u+1) \sqrt{2 u+1} < (u+2)^2.
\end{equation}
Indeed, $(u+2)^2-2 (u+1) \sqrt{2 u+1}  \\
=2+\left(u+1-\sqrt{2 u+1}\right)^2 > 0$.
This ends the proof of the inequality \eqref{Append_0}, 
or, equivalently, $y(p,a) < 1$ for all positive $p$ and $0 < a < 1$.



\end{document}